\documentclass[12pt,reqno]{amsart}

\usepackage{geometry} % see geometry.pdf on how to lay out the page. There's lots.
\usepackage{graphicx}% Include figure files

\usepackage[scaled=0.92]{helvet}	% set Helvetica as the sans-serif font

\usepackage{xr} %external references
\def\T{T}

\theoremstyle{definition}

\usepackage{listings}
\usepackage{color} %red, green, blue, yellow, cyan, magenta, black, white
\definecolor{mygreen}{RGB}{28,172,0} % color values Red, Green, Blue
\definecolor{mylilas}{RGB}{170,55,241}
\lstdefinestyle{myCustomMatlabStyle}{
  language=Matlab,
  numbers=left,
  stepnumber=1,
  numbersep=10pt,
  tabsize=4,
  showspaces=false,
  showstringspaces=false
}

\usepackage[table]{xcolor}
 \definecolor{darkblue}{rgb}{0.0, 0.0, 0.55}
 \definecolor{brightmaroon}{rgb}{0.76, 0.13, 0.28}

\usepackage{mathrsfs} 
\usepackage{dsfont}
\usepackage{hyperref}
\hypersetup{
    colorlinks = true,
    linkbordercolor = {white},
    citecolor = darkblue,
    linkcolor = brightmaroon,
}

\usepackage{etoolbox}
\usepackage{amsaddr}
\makeatletter
\patchcmd{\@setauthors}{\uppercasenonmath\@setauthors}{}{}{}
\makeatother

\usepackage[numbers,sort&compress,square]{natbib}

\newcommand{\be}{\begin{equation}}
\newcommand{\ee}{\end{equation}}
\newcommand{\bea}{\begin{eqnarray}}
\newcommand{\eea}{\end{eqnarray}}
\newcommand{\bw}{\begin{widetext}}
\newcommand{\ew}{\end{widetext}}

\newcommand{\bi}{\begin{itemize}}
\newcommand{\ei}{\end{itemize}}

\renewcommand{\d}{\ensuremath{\operatorname{d}\!}}
\newcommand{\matr}[1]{\mathbf{#1}}

\usepackage{bm}

\title[Sensitivity of Complex Networks]{Sensitivity of Complex Networks}

\date{\today} % delete this line to display the current date

\author{\small Marco Tulio Angulo$^{\MakeLowercase{1,2,6}}$,  \ Gabor Lippner$^\MakeLowercase{3}$,  Yang-Yu Liu$^{\MakeLowercase{2,4}}$, Albert-L\'{a}szl\'{o} Barab\'{a}si$^{\MakeLowercase{1,4,5}}$}
\address[Author 1]{
$^1$Center for Complex Networks Research, Northeastern University,
Boston MA 02115, USA\\
$^2$Channing Division of Network Medicine, Brigham and Women's
Hospital, and Harvard Medical School, Boston MA 02115, USA\\
$^3$Department of Mathematics, Northeastern University, Boston MA 02115, USA.\\
$^4$Center for Cancer Systems Biology, Dana-Farber Cancer Institute,
Boston MA 02115, USA\\
$^5$Center for Network Science, Central European University, Budapest 1052, Hungary. \\
$^6$ Present address: CONACyT Research Fellow at the Institute of Mathematics, Universidad Nacional Aut\'onoma de M\'exico (UNAM), Juriquilla 76230, M\'exico.
 \vspace*{2cm}
}
\begin{document}

\begin{abstract}
The sensitivity (i.e. dynamic response) of complex networked systems has not been well understood, making difficult to predict whether new macroscopic dynamic behavior will emerge even if we know exactly how individual nodes behave and how they are coupled. 
Here we build a framework to quantify the sensitivity of complex networked system of coupled dynamic units. 
We characterize necessary and sufficient conditions for the emergence of new macroscopic dynamic behavior in the thermodynamic limit. 
We prove that these conditions are satisfied only for architectures with power-law degree distributions.
Surprisingly, we find that highly connected nodes (i.e. hubs) only dominate the sensitivity of the network up to certain critical frequency. 
 
\end{abstract}

\maketitle
%\tableofcontents

\newpage

\section{Introduction}
%%%%%%%%%%% network science: interplay betweeen structure and dynamics

Understanding how macroscopic dynamic behavior can emerge from a networked system of many agents (or units) is  a fundamental question in  physics     \cite{anderson1972more,gu2009more}. Addressing this question also has  implications in  biology \cite{amaral2004emergence}, engineering \cite{niazi2011sensing} and cognitive science \cite{thompson2001radical}. 
Complex  behavior can emerge purely from nonlinear dynamics ---through bifurcations \cite{sparrow2012lorenz}, catastrophes \cite{arnol1992catastrophe}, strange attractors \cite{strogatz2014nonlinear} and so on--- but may also emerge due to the aggregation of a large number of agents with simple nodal dynamics in  the so-called \emph{thermodynamic limit}.  A famous example is Boltzmann's $H$-theorem, where agents with  time-reversible nodal dynamics give rise to macroscopic irreversible behavior  \cite{villani2008h}.
However, it remains unclear how and when the properties of this aggregation ---given by the network describing which agent interacts with whom--- will lead to the emergence of  new macroscopic dynamic behavior.%
%Is it possible that new behavior emerges

Here we introduce a framework to study the \emph{sensitivity} (i.e. dynamic response) of networked systems and investigate the conditions for the emergence of new macroscopic behavior.
%from a network perspective. 
%
%The framework allows quantifying the dynamic response of each agent in the system, and makes mathematically precise the notion of ``emergence of new behavior".
%Here we consider multi-agent systems describing $N$ interacting nodes,  we introduce a framework to study  their dynamic response form a network perspective and  that also makes mathematically precise the notion of ``emergence of new behavior''. 
%
We first analyze how the degree of a node in the network shape its contribution to the dynamic response of the {system}.
We find that in systems with high-order nodal dynamics, the hubs (i.e., high-degree nodes)  not always dominate  the dynamic response of the {system}.
In other words,  nodes important from a network perspective are not as important from a dynamic perspective. 
Indeed, with second-order oscillatory dynamics, we find there is a transition point close to the resonant frequency of the dynamics where  hubs lose their dominant role.
Then we study how the interconnection topology of a system (i.e., the network topology) shapes its macroscopic  or collective behavior, finding that the degree distribution of the network is sufficient to constraint the emergence of new behavior. 
We rigorously prove that new behavior emerges if and only if no eigenvector of the interconnection network aligns with the vector $(1, \cdots, 1)^T$. In other words, new behavior emerges if and only if the system cannot fully synchronize.
In particular, we show that new behavior cannot emerge in the thermodynamic limit using interconnection networks with Erdos-Renyi architecture, in the sense that their dynamic behavior can be reproduced by the dynamics of a single node. 
 In contrast, we prove that new behavior emerges with degree distributions with heavy tails, e.g., power-law degree distributions in scale-free networks.

%%%%%%%%%%%%
%%%%%%
%%%%%%
%%%%%%
%%%%%%%%%%%%

\section{Model}

In order to focus on the role of the aggregation, we assume that each node (agent) has a simple linear dynamics. Under this assumption, a broad class of dynamic systems describing $N$ interacting agents in an undirected network $\mathcal G$ can be described by
\begin{equation}
\label{system}
D {\bm x}(t) =\matr{A} {\bm x}(t),% \quad {\bm x}(0)={\bm x}_0,%+ B {\bm v}
\end{equation}
where  ${\bm x} = (x_1, \cdots, x_N)^\T$ with $x_i$  the activity of node $i$.
%and ${\bm v} = (v_1, \cdots, v_m)$ is a vector of exogenous inputs (e.g., control inputs). 
%
The network $\mathcal G$ underlying the system is encoded by the symmetric interaction matrix $\matr A = (a_{ij}) \in \mathbb R^{N \times N}$, representing the direct interactions between the agents:  $a_{ij} \neq 0$ if nodes $i$ and $j$ directly interact (i.e., $\mathcal G$ contains a link between nodes $i$ and $j$ ) and $a_{ij} = 0$ otherwise. 
The nonzero edge-weights are given by  $a_{ij} = \frac{1}{\kappa} \rho_{ij}$
where $\rho_{ij}>0$ represents the interaction strength between agents $i$ and $j$. 
We assume the $\rho_{ij}$'s follow a fixed distribution $\rho$ that we choose over the interval $(0,1]$ without loss of generality.
% This number is  chosen randomly from a uniform distribution and, due to the linearity of the system dynamics, we can consider this distribution to be uniform over the interval $[0,1]$ without loss of generality. 
The mean degree  $\kappa$ of the network acts as a scaling factor that is  necessary to obtain a bounded mathematical object when we take the  thermodynamic limit $N \rightarrow \infty$.
Dynamics enter  \eqref{system} through $D=g \matr{I}_{N \times N}$, where $g$ is a linear causal operator  acting on the trajectory of each node $x_i(t)$ and  $\matr{I}_{N \times N}$ is the identity matrix of dimension $N$. For example, if $g$ is the derivative operator $\d/\d t$ then \eqref{system} is the first-order system $\dot{\bm  x}(t) = \matr{A} {\bm x}(t)$ usually found in diffusion  or consensus \cite{delvenne2015diffusion}.  In the case $g= (\d/\d t)^2 + 2 \zeta \omega_n \zeta (\d/\d t) + \omega_n^2$ then \eqref{system} becomes $\ddot {\bm x}(t) + 2 \zeta \omega_n {\bm x}(t)+ \omega_n^2 {\bm x}(t) = \matr{A} {\bm x}(t)$, which represents $N$ coupled oscillators with damping $\zeta$ and natural frequency $\omega_n$ often used for modeling the power-grid  \cite{kundur1994power}.

It is convenient to rewrite system \eqref{system} in the Laplace domain 
\begin{equation}
\label{system-laplace}
D {\bm x}(s) =\matr{A} {\bm x}(s) + {\bm v}(s, {\bm x}(0), \dot {\bm x}(0), \dots  )%+ B {\bm v}
\end{equation}
where  ${\bm x}(s) = \int_0^\infty  {\bm x}(t) e^{-st} \d t$ is the Laplace transform of ${\bm x}(t)$, $s \in \mathbb C$ is the Laplace variable, and ${\bm v}$ is a vector of initial conditions of the system\footnote{Recall that in Laplace domain, the  derivative operator $\d/\d t$ is given by multiplication by $s$. Therefore, when $g=\d/\d t$ then $g(s) = s$ and ${\bm v}(s,  {\bm x}(0)) =  {\bm x}(0)$. But if $g= (\d/\d t)^2 + 2 \zeta \omega_n \zeta (\d/\d t) + \omega_n^2$ then $g(s)=s^2 + 2 \zeta \omega_n s + \omega_n^2$ and ${\bm v}(s,{\bm x}(0),\dot  {\bm x}(0)) = (s + 2 \zeta \omega_n)  {\bm x}(0) + \dot  {\bm x}(0)$ }.  Then, equation  \eqref{system-laplace} can be rewritten as
$${\bm x}(s) = \matr{S}(s) {\bm v}(s, {\bm x}(0), \dot {\bm x}(0), \dots  ) $$
where  $\matr{S}(s) = (D(s) -\matr{A})^{-1} = (g(s)\matr{I} - \matr{A})^{-1} \in \mathbb C^{N \times N}$ maps initial conditions to trajectories, and thus is known as the \emph{transfer} or \emph{sensitivity} function of the system \cite{astrom2010feedback}.
 
%%%%%%%%%%%%%%%%%%%%%%%%
%%%%%%%%%%%%
%%%%%%%%%%%%
%%%%%%%%%%%%
%%%%%%%%%%%%
%%%%%%%%%%%%%%%%%%%%%%%%

\section{Results}

%\subsection{Node-level sensitivity}
To obtain the macroscopic behavior of the system, assume that all agents start from the same initial condition ${\bm x}(0) = \mathds 1 x_0$, $\dot {\bm x}(0) = \mathds 1 \dot x_0$, $\dots$, where $\mathds 1 = (1, \cdots, 1)^\T$. Consequently, ${\bm v} = \mathds 1 v(s, x_0, \dot x_0, \dots)$ and the dynamic behavior of each agent in the system is ${\bm x}(s) = {\bm S}_N(s) v $, where
$$ {\bm S}_N(s) =  \matr{S}(s) {\mathds 1}$$
is an $N$-dimensional vector containing the \emph{node sensitivity}. This vector characterizes the dynamic response of each agent of the system.  
  Similarly, the average or macroscopic behavior of the system $\bar x = N^{-1} \sum_i x_i$ is given by  $\bar {x}(s) = \bar S_N(s)v$,  where
$$\bar S_N(s) =  \frac{1}{N}  {\mathds 1}^T (g(s) \matr{I} -\matr{A})^{-1} {\mathds 1}= \frac{1}{N}  {\mathds 1}^T \matr{S}(s) {\mathds 1} $$
is the \emph{mean sensitivity} or  mean transfer function of the system, characterizing its average dynamic behavior. 
With this notation the dynamic behavior of the isolated agents (i.e. $\matr A = \matr 0$) is $f(s) = 1/g(s)$  and the dynamic behavior \textcolor{red}{of the system} in the thermodynamic limit is $\bar S_\infty (s) = \lim_{N \rightarrow \infty} \bar S_N(s)$. %  (with possibly different parameters e.g. different damping and natural frequency if $g$ is an oscillator).  

\begin{figure*}[!t]
\centering
\includegraphics[width=7.4in]{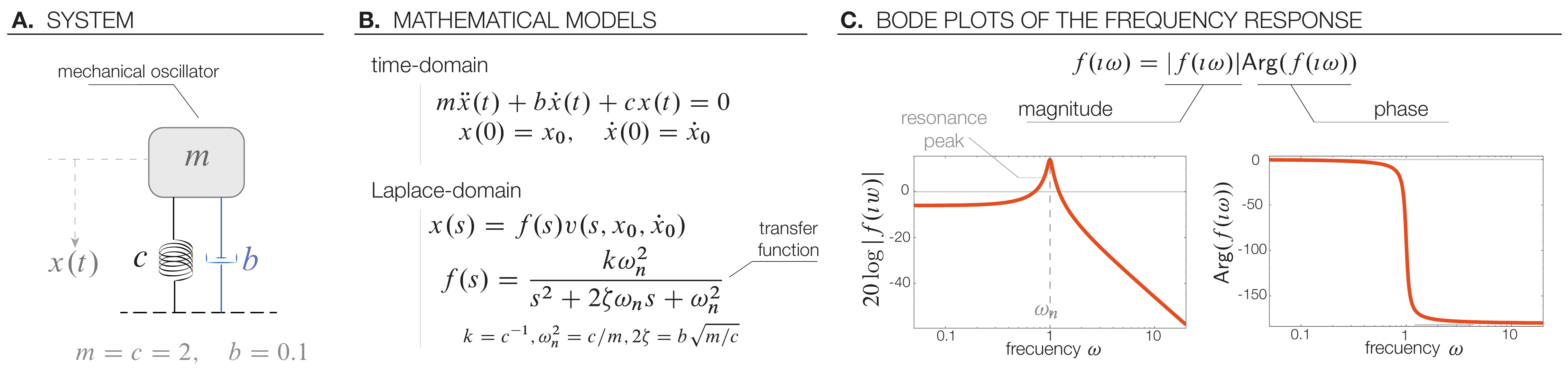}
\caption{{\bf The dynamics of a system is characterized by the frequency response of its sensitivity function.}
 {\bf A.} A spring-mass-damper system (with parameters $c,m,b$ ) is a mechanical oscillator where $x(t)$ is the position of the mass
  {\bf B.} The system can be modeled in time domain using an ordinary differential equation, or in Laplace domain using a transfer function. 
  {\bf C.} A transfer function $f(s)$ can be equivalently written in the Fourier domain $f(\imath \omega)$ using its frequency response  $s= \imath \omega$ with $\omega \in [0, \infty)$. The frequency response  characterizes the system's dynamics describing what are the amplitude gain $|f(\imath \omega)|$ and phase lag ${\sf Arg}(f(\imath \omega))$ that the system would exhibit if a sinusoidal input with frequency $\omega$ were applied as  input $v$. Here we observe the well-known resonance peak of an oscillator at its natural frequency $\omega_n$. The Bode plots show the magnitude (in decibels $20 \log(\cdot)$) and phase of the transfer function.}
\label{fig:freq-resp}
\end{figure*}

With the above framework, new behavior emerges when $\bar S_\infty(s)$ cannot be approximated by $f(s)$.
 One method to quantify the difference between   $f(s)$ and $\bar S_\infty(s)$ is comparing their \emph{frequency responses} or Fourier transforms. The frequency response is obtained from the Laplace transform by substituting $s = \imath \omega, \imath^2=-1$, and letting the frequency $\omega$ vary from $0$ to $\infty$. Then, $f(s)$ is near $\bar S_\infty(s)$  if the complex number $f(\imath \omega)$ is near $\bar S_\infty(\imath \omega)$ for $\omega \in [0, \infty)$. We can graphically read this more conveniently using Bode plots of magnitude and phase, Fig. \ref{fig:freq-resp}.

%%%%%

\subsection{Node-level sensitivity} 
\begin{figure*}[!t]
\centering
\includegraphics[width=7.4in]{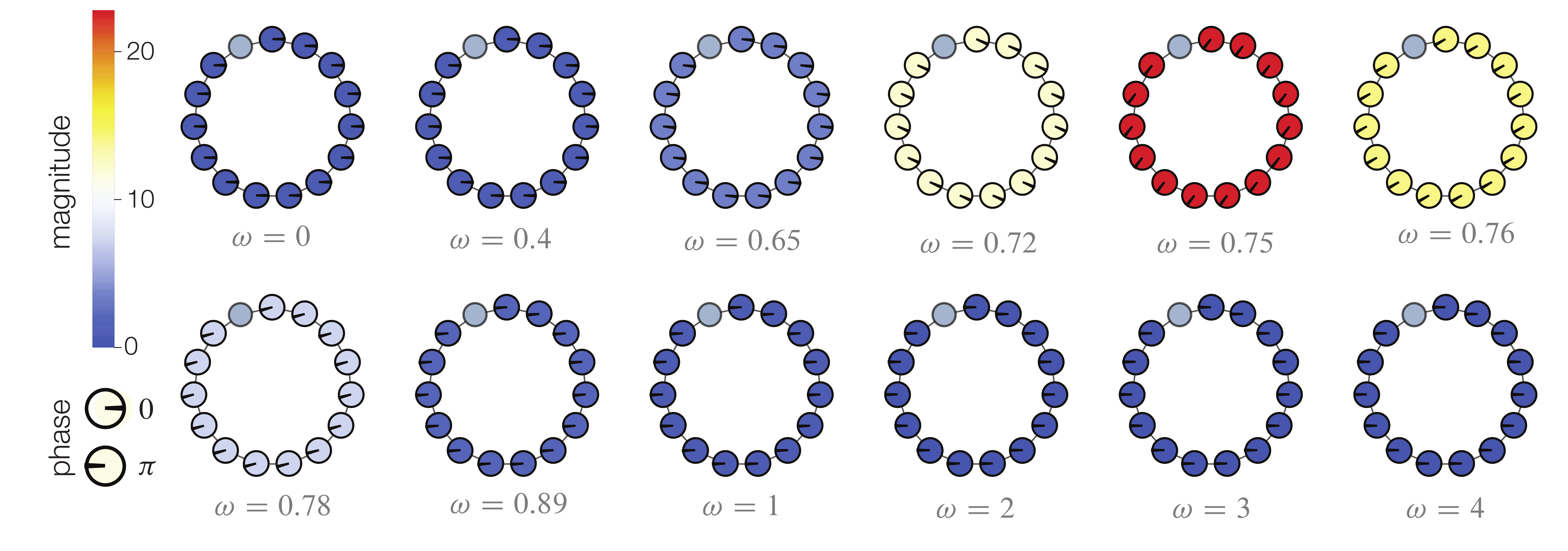}
\includegraphics[width=7.4in]{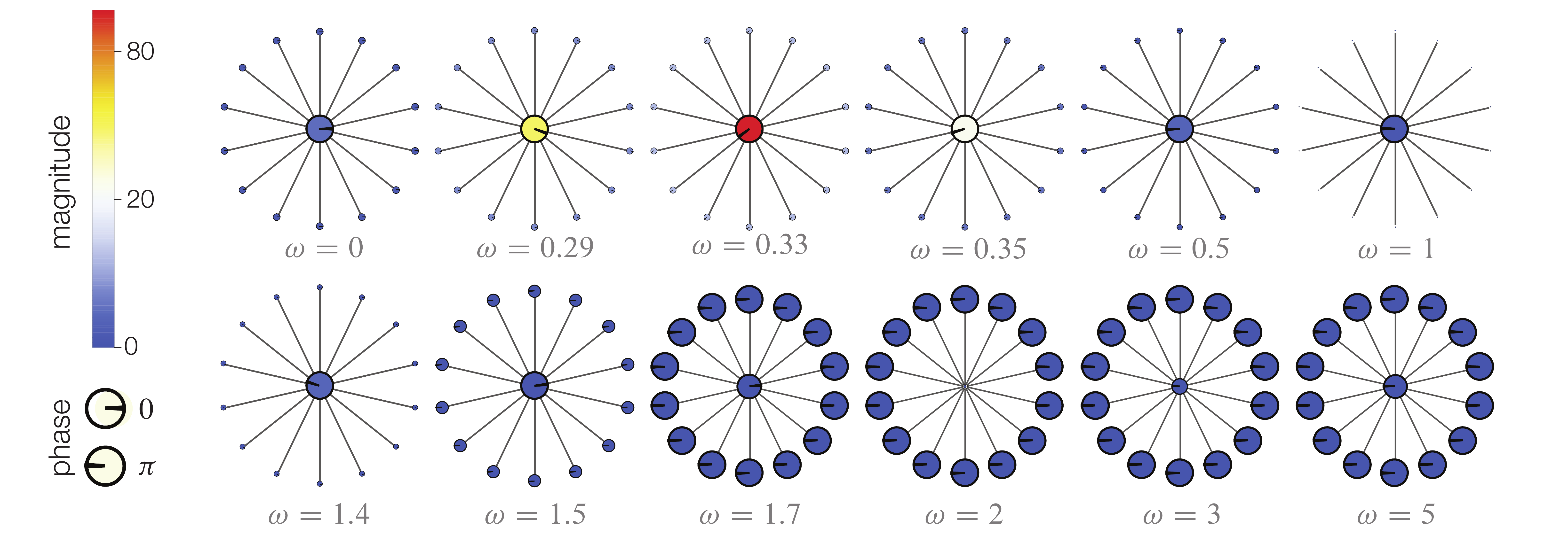}
\includegraphics[width=7.4in]{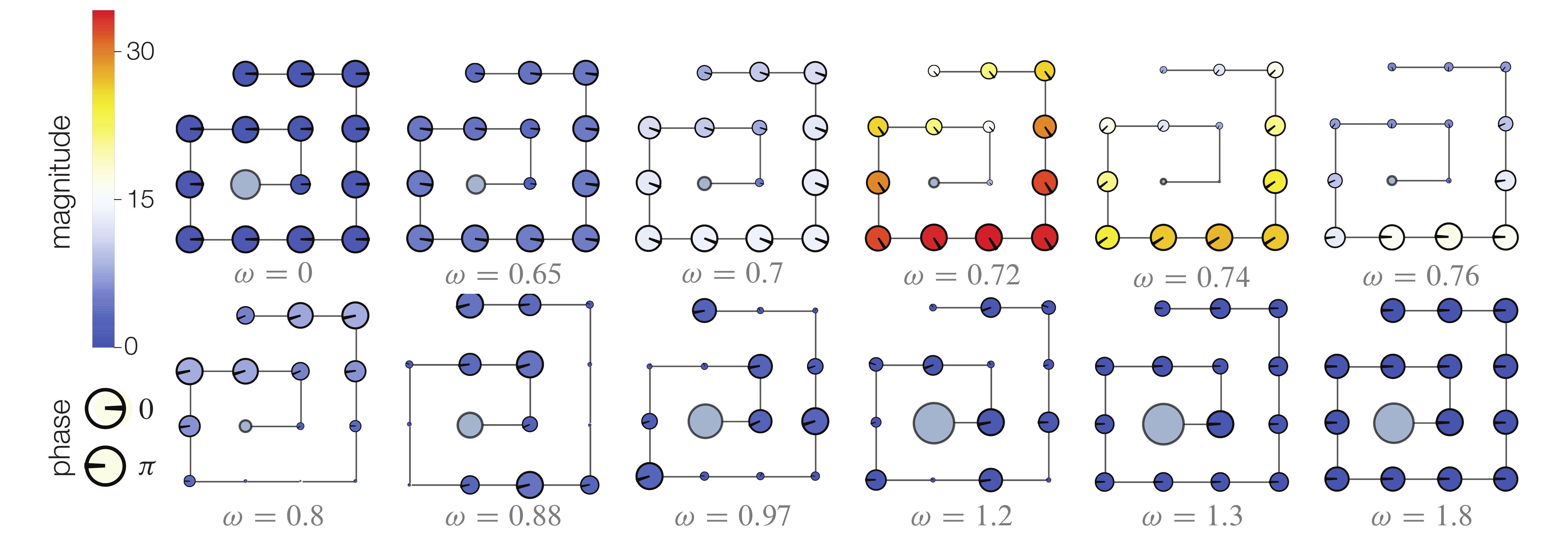}
\caption{{\bf Node level sensitivity.} The dynamics are of second order with $\omega_n =1$, $\zeta=0.01$ and $k=(1-0.1)/\max_i \lambda_i$, where $\lambda_i$ runs over the eigenvalues of all networks ensuring that the system is stable. The edge-weights of the network are chosen as $1$.  }
\label{fig:nodelevel-sens}
\end{figure*}

\begin{figure*}[!t]
\centering
\includegraphics[width=7.4in]{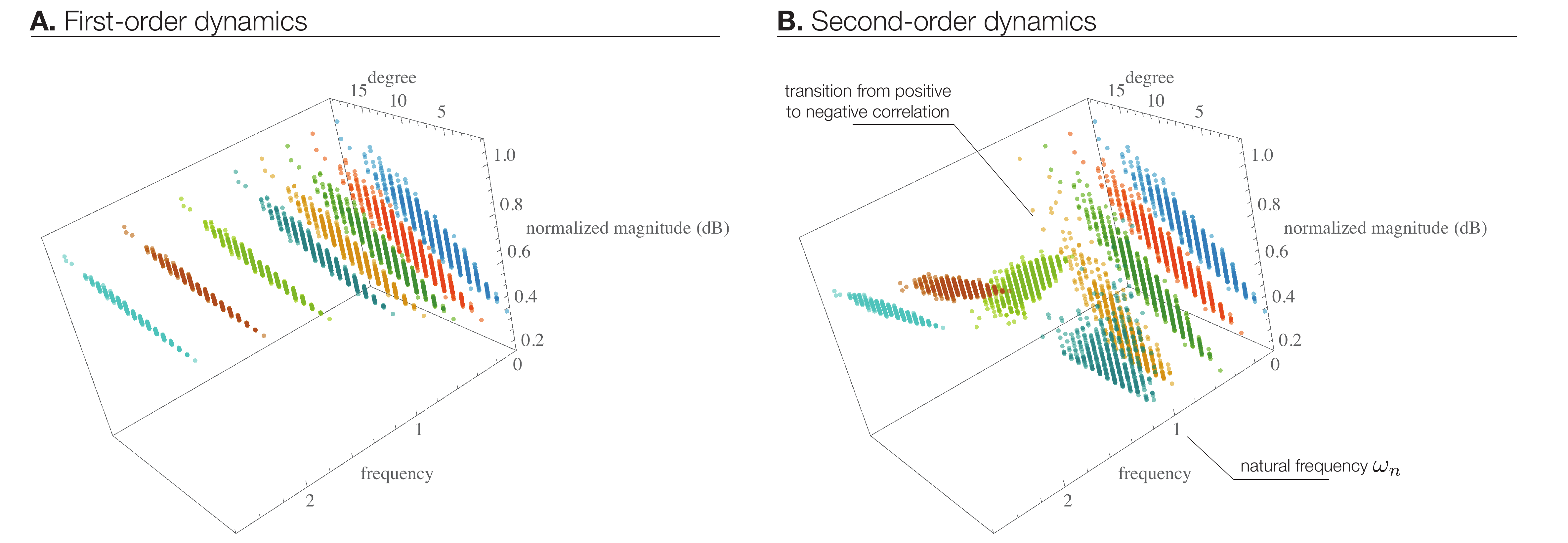}
\includegraphics[width=7.4in]{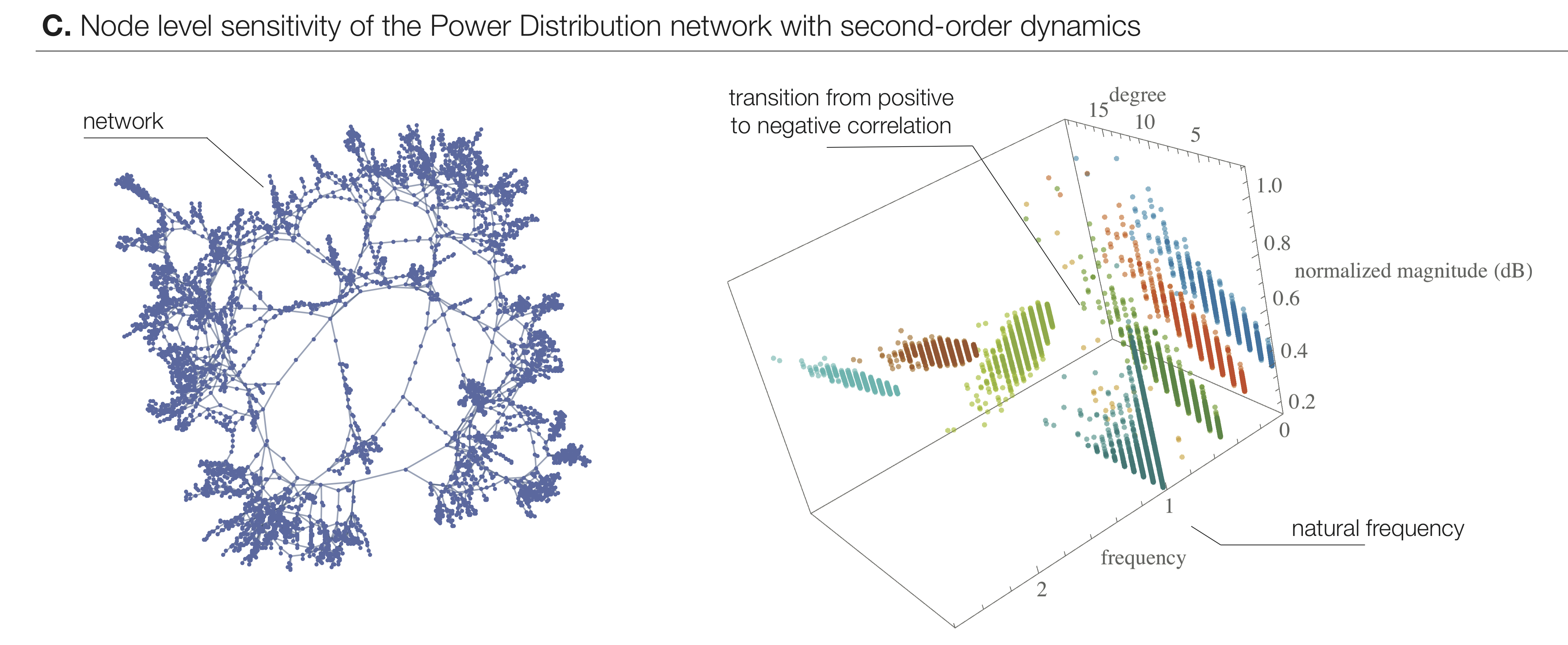}
\caption{{\bf Correlation between node degree and magnitude of its sensitivity.}  {\bf A.} For first-order dynamics $g(s) = (s + \omega_n^2)/ (k \omega_n^2)$, the larger the degree of the node the larger the magnitude of its response. Here we used the parameters $\omega_n=1$ and $k=0.5(1-c)/ \max_i \lambda_i(A)$, $c=0.1$ ensuring that the system is stable.  The  $\matr A$ matrix is built from an Erdos-Renyi architecture with edge-weights randomly chosen from a uniform distribution on $[0,1]$. 
 {\bf B.} For second-order dynamics $g(s) = (s^2 + 2 \zeta \omega_n s + \omega_n^2)/ (k \omega_n^2)$ there is a transition point close to the natural frequency of the system $\omega_n$: for frequencies $\omega < \omega_n$ nodes with large degree tend to have large magnitude of response; for $\omega > \omega_n$ nodes with low degree tend to have larger magnitude of response. Here we used the same parameters $\zeta=0.01$, with the other parameters (including the network $\matr A$) the same as with first-order dynamics. 
 {\bf C.}  Degree versus node-level sensitivity magnitude for the Power Distribution network with second-order dynamics $N=4591$.   }
\label{fig:nodelevel-sens2}
\end{figure*}

The node level sensitivity $\bm S_N(s)$  characterizes the nodes that are important from  a dynamical viewpoint, assuming all agents start from the same initial condition.
In this case, the dynamic response of node $i$ at frequency $\omega$ is the $i$-th element of the vector $\bm S_N(\imath \omega)$. Naturally, the elements of $\bm S_N(\imath \omega)$ with larger magnitude contribute more to the average behavior of the system at frequency $\omega$.
As shown in Fig. \ref{fig:nodelevel-sens},   for second-order dynamics, the hubs  dot not always have the response with largest magnitude. For example,  in the star network, the central hub has a larger magnitude than the leaf nodes at low frequencies, but smaller magnitude at high frequencies.  
Indeed, we observe that the leaf nodes are always synchronized (i.e., have equal  phase), but they are not always synchronized with the hub. 
Since the response of each node is a weighted sum of its nearest neighbors (note that this is a sum of complex numbers), the contribution of the hub is maximal at $\omega =1$ because the leaves and hub synchronize.
In contrast, at $\omega=2$, the  hub and the leaves are anti-synchronized (i.e., have opposite phases) and now the leaves inhibit the response of the hub.
We found  a similar phenomenon  in larger networks  with second order dynamics  ---using networks either randomly generated or real ones such as a power grid--- irrespectively of the distribution of the edge-weights, Fig.\ref{fig:nodelevel-sens2}. 
Nodes in a power grid correspond to generators, and a standard model for their dynamics is the so-called ``swing dynamics'', which is linear and of second order \cite{kundur1994power}. 
When the dynamics are of first order, this phenomenon disappears. 
Real systems always contain high-order dynamics ---in principle, any system  can be modeled with  arbitrary high precision by a linear model with sufficiently high order--- and hence this result shows that the topological properties of the interconnection network do not determine completely the nodes that dominate the response of the system at all frequencies. 
In other words, if we are interested in understanding the low-frequency dynamics of a  networked system it is enough to focus on the most connected agents such as the hubs. 
However, in order to understand the high-frequency dynamics of a  networked system, we must  consider the agents which are  in the ``periphery'' of the network. % and that are harder to map \textcolor{red}{(e.g., the fewer connections a gene has in a gene regulatory network, the harder is to find and perturb genes that interacts with it and be able to map it)}.
This counterintuitive effect of weakly connected nodes illustrates the rich interplay between networks and dynamics in complex systems.

%%%%

\subsection{Network-level sensitivity} The network-level sensitivity characterizing the macroscopic dynamic behavior of the system depends on the network encoded in the interaction matrix $\matr A$. In principle, one expects that  networks with different edge-weights or interconnection topologies (e.g., Erdos-Renyi or scale-free topologies)  produce different mean sensitivities (Fig.\ref{fig:nodelevel-sens}). 
In what follows we show this is not the case in the thermodynamic limit $\bar S_\infty(s)$, and that  there are  two ``sensitivity classes''  ---in the first, the mean behavior of the system equals the behavior of the isolated nodes, and in the second new behavior emerges--- that depend on  the heterogeneity of connections of the network only.
%
%More precisely, regardless of the specific edge-weights $a_{ij}$'s of $\matr A$, very coarse (and macroscopic) properties of the interconnection network $\mathcal G$ such as its degree distribution can constraint importantly the possible thermodynamic limit $\bar S_\infty(s)$.

First rewrite $\bar S_N(\imath \omega)$ in terms of the eigenvalues $\{\lambda_i\}_{i=1}^N, \lambda_1 \geq \cdots \geq \lambda_N$, and eigenvectors $\{\bm \varphi_i\}_{i=1}^N$ of $\matr A$, yielding
\begin{equation}
\label{sum-sensitivity}
\bar S_N(\imath \omega) = \sum_{i=1}^N  h_i ( \imath \omega)  \frac{\langle \mathds 1, {\bm \varphi}_i \rangle ^2}{N}, \quad .
\end{equation}
where $h_i(s)=f(s)/(1-\lambda_i f(s))$. This equation shows that the average response of the system is a weighted sum of the functions $h_i(s)$. 
Hence new behavior emerges if and only if no eigenvector  aligns with the vector $\mathds 1$.
Since $\matr A$ is symmetric, the eigenvectors  $\bm \varphi_i \in \mathbb R^N$ can be chosen real, orthogonal and with unit norm, so they satisfy
\begin{equation}
\label{sum-weights}
\langle \mathds 1, \bm \varphi_1 \rangle ^2/ N +  R  = 1, 
\end{equation}
where $R=(\langle \mathds 1, \bm \varphi_2 \rangle ^2 + \cdots +  \langle \mathds 1, \bm \varphi_N \rangle ^2)/N$. 

 From  equations \eqref{sum-sensitivity} and \eqref{sum-weights}, we conclude  that $\bar S_\infty (s)$  can be approximated by $f(s)$ if and only if one term in the sum \eqref{sum-weights} dominates as $N \rightarrow \infty$. We can prove that when the interconnection network is an Erd\"os-Renyi (ER) network then $\bm\varphi_1 \rightarrow \mathds 1$ as $N \rightarrow \infty$ (Theorem 1 in SI). Therefore,  the weight associated to the first eigenvalue is dominant 
$$  \langle \mathds 1, \bm \varphi_1 \rangle ^2/N = 1 - O(1/\sqrt{pN}), \quad N \rightarrow \infty, $$
where $p \gg (\log^4 N)/N$ is the probability of connection in the ER random network model. This implies that  the contribution of the residue  $R = O(1/\sqrt{p N})$ vanishes in the thermodynamic limit, and that $\lambda_1 = 1 + O(1/\sqrt{p N})$.  Consequently, assuming the system is stable, these two results  imply that
$$\bar S_\infty(s) = \frac{f(s)}{1-f(s)} $$
showing that new dynamic behavior cannot emerge in ER networks.
 For example, considering $f(s)$ to be the oscillator dynamics $(k \omega_n^2) / (s^2 + 2 \zeta \omega_n s + \omega_n^2)$ then we have that
$$\bar S_\infty(s) = \frac{k_{\sf ER} \omega_{n,\sf ER}^2}{s^2 + 2 \zeta_{\sf ER} \omega_{n,\sf ER} s + \omega_{n,\sf ER}^2} $$
 is again an oscillator with parameters  $\omega_{n, \sf ER}= \omega_n \sqrt{1 - k}$, $\zeta_{\sf ER}= \zeta/\sqrt{1-k}$ and $k_{\sf ER}= k/(1-k)$. 
Numerical experiments show that  several other interconnection topologies such as lattices, small-world or random geometric networks also belong to this first  class in which new macroscopic behavior do not emerge (Fig. 5).
  Note  also that, due to the law of large numbers,  the thermodynamic limit of a network without structure (i.e., $\mathcal G$ is the complete graph on $N$ nodes)  also belongs this first  class.

\begin{figure*}[!t]
\centering
\includegraphics[width=7.4in]{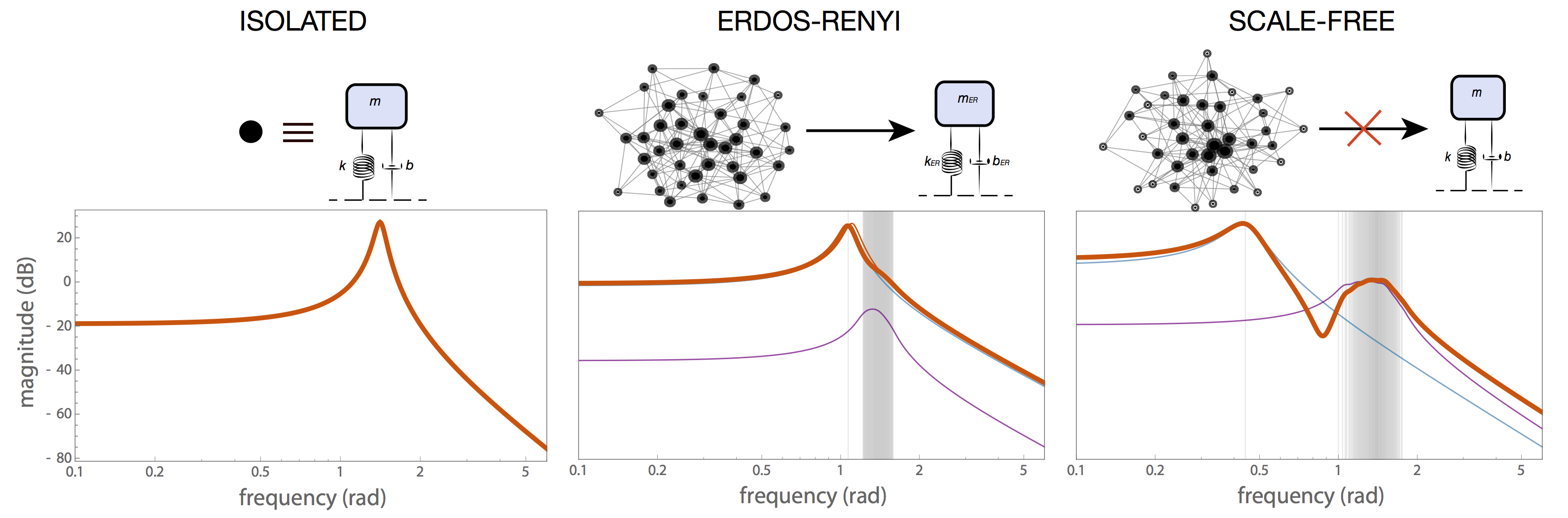}
\put(-260,135){\fontsize{5}{5pt}mean}
\put(-265,123){\fontsize{8}{10pt} $N \rightarrow \infty$}
\put(-89,120){\fontsize{8}{10pt} $N \rightarrow \infty$}
\put(-80,140){\fontsize{5}{5pt}mean}
\caption{{\bf Network-level sensitivity in the thermodynamic limit.} The mean sensitivity function in ER networks converges to that of a scalar node, but with different parameters (i.e. $f/(1-f)$). In contrast, the mean sensitivity of SF networks can not be realized by the transfer function of a single component, giving rise to new behavior. 
The response due to the first eigenvalue is shown in blue, and the response due to the rest of eigenvalues (i.e., $R$) is shown in purple. 
In the ER case the response of the first eigenvalue dominates, while in the SF case the response due the  rest of eigenvalues dominates at high frequency.
We illustrate this using the transfer function of a simple harmonic oscillator and $N=2048$ nodes. \emph{Left:} frequency response of the isolated system with $\omega_n=\sqrt{2}, {\zeta}=0.05$ and $\epsilon =0.1$. Its gain is selected as $k=0.37949$ to ensure that the interconnected system remains stable using both networks. \emph{Middle}: using  $p=0.005$, we generate a Erdos-Renyi network which has $10411$ edges and $\bar k=10.167$. \emph{Right}: using the $m=5$, we generate a Barabasi-Albert network with $10225$ edges and $\bar k= 9.985$.}
\label{fig:Fig1}
\end{figure*}

The situation is drastically different in networks with heavy-tailed degree distribution ${\sf Pr}(\deg(v) = k) \approx (\gamma-1)k^{-\gamma}$. In Theorem 2 of SI, we proved that if $\gamma>2$ the contribution due to the first eigenvalue satisfies
$$  \langle \mathds 1, \bm\varphi_1 \rangle ^2/N = O(1/N^\beta), $$
where $\beta =1/(\gamma-1)$. This implies that $R=1-O(1/N^\beta)$ and the dynamic behavior in thermodynamic limit is  opposite to ER networks: the response due to the first  term vanishes as $N \rightarrow \infty$ and the residue $R$ dominates. Furthermore, the weight of all other eigenvectors remain approximately the same. 
This fundamental distinction of networks with  heavy-tailed degree distribution becomes evident by comparing their frequency responses, Fig. \ref{fig:Fig1}. % Indeed,  numerical evidence suggests that new behavior does not emerge in lattice, small-world or (random) geometric networks since in all of them $\mathds 1$ asymptotically becomes an eigenvector (Fig. \ref{fig:Fig2}). 
Only in these networks do we observe two peaks in their frequency response that cannot be approximated by the response of an isolated node (which has a single bump corresponding to the resonance peak). 
Thus, the thermodynamic limit of SF networks belong to a second sensitivity class in which new macroscopic behavior emerges.

\begin{figure*}[!t]
\centering
\includegraphics[width=7in]{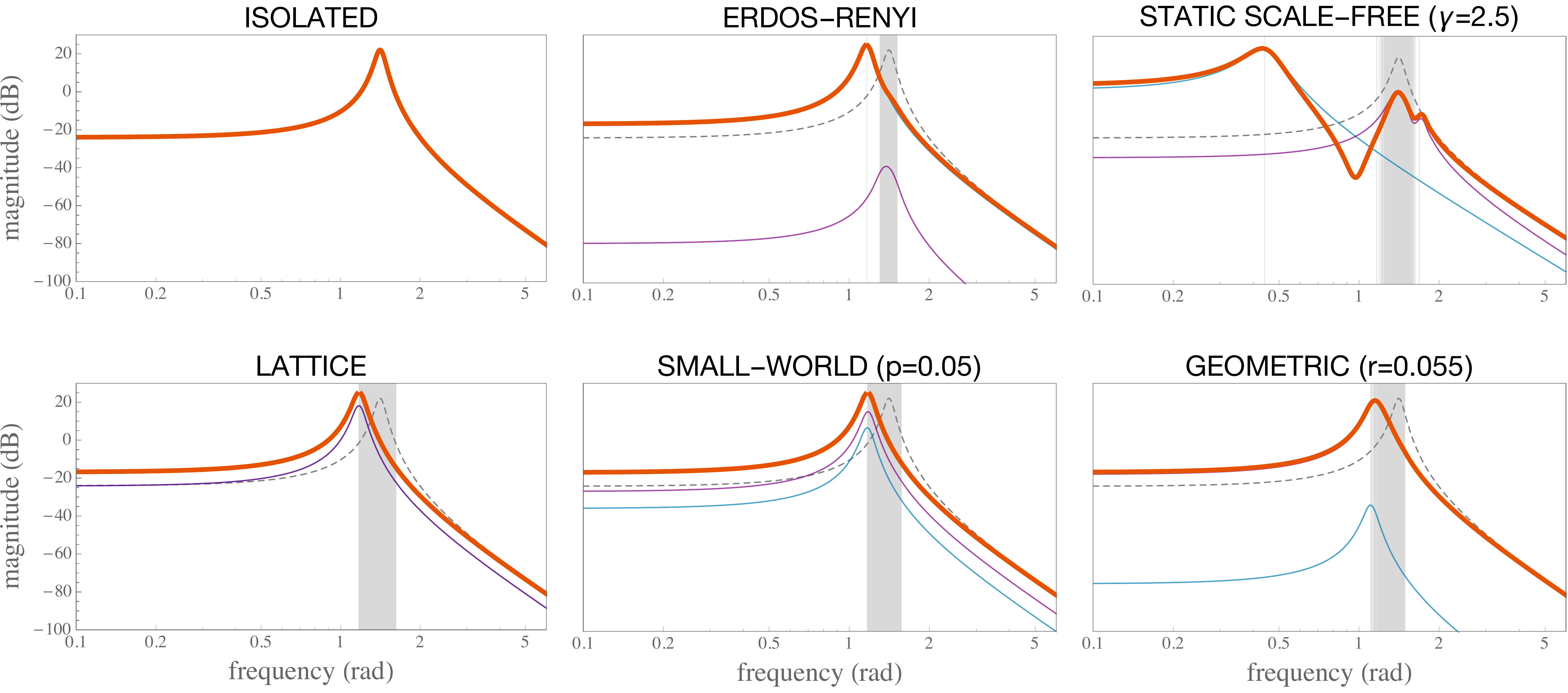}
\caption{{\bf Network-level sensitivity in the thermodynamic limit for different interconnection architectures.}  Here we approximate $S_\infty( \imath \omega)$ by taking $S_N( \imath \omega)$ for $N=2048$, and use second-order dynamics with parameters as in Fig. \ref{fig:Fig1}.
The response due to the first eigenvalue is shown in blue, and the response due to the rest of eigenvalues (i.e., $R$) is shown in purple. 
New behavior does not emerge for Erdos-Renyi, Lattice, Small-World (Watts-Strogatz) or (random) Geometric architectures. Only for scale-free networks we observe a second bump that can not be approximated by the response an isolated node. 
}
\label{fig:Fig2}
\end{figure*}

%%%%%%%%%%%%%%%%%%%%%%%%
%%%%%%%%%%%%
%%%%%%%%%%%%
%%%%%%%%%%%%
%%%%%%%%%%%%
%%%%%%%%%%%%%%%%%%%%%%%%

\section{Outlook and concluding remarks}

Large systems can exhibit complex   behavior due to their dynamics and/or due to the properties of its interconnection network.
By focusing on linear dynamics, we have shown that very coarse properties of the interconnection network such as  its degree sequence can determine the macroscopic behavior of the system. 
%We have shown that very coarse properties of the interconnection network of a system sucha
% We have characterized how  
%interconnection properties such as node degree can  influence the dynamics of the networked system.
%
We found that homogenous interconnections (as cycles) tend to produce 
similar dynamic response in all nodes in the network. 
But  heterogeneous interconnections (like stars or even paths) cause 
that each node contributes differently to the response and the system, with magnitude and phase depending on the specific frequency that is excited. 
%
%For high-order dynamics,  nodes that are important in the network (i.e., those with large degree) not always have a large contribution from the dynamic perspective.%, a phenomenon that is caused by second and higher-order dynamics.
%
We  identified the mechanism through  which the interconnection network favors the emergence of new behavior in the thermodynamic limit.  New behavior emerges when the eigenvectors of the interconnection network do not align with the vector $\mathds 1$, which represents consensus.  
Furthermore, we showed there are two sensitivity classes, one in which new behavior does not emerge ---containing unstructured, ER and several other topologies---  and the other in which new behavior emerges ---containing SF networks.
A similar analysis for the case of directed graphs remains open, mainly due to the fact that the spectral theory of random directed graphs is not as  developed as in the case of undirected graphs. 
Also, a better understanding of how  nonlinearities and heterogeneities in the node dynamics may affect  wether the system belongs to the first or second sensitivity class can provide further insights into the role of the interconnection network in the emergence of macroscopic behavior.

% Our results also mean field theories. 

%{\it Author contributions:} All authors designed and did the research. M.T.A. and G.L. did the analytical analysis. M.T.A. and Y.-Y. L. wrote the manuscript. A.L.B. edited the manuscript. 

%{\it Author information:} Correspondence and requests for materials should be addressed to Y.-Y.L. (yyl@channing.harvard.edu).

%{\it Acknowledgements:} This work was supported in part by the John Templeton Foundation.

%%%%%%%%%%%%%%%%%%%%%%%%
%%%%%%%%%%%%
%%%%%%%%%%%%
%%%%%%%%%%%%
%%%%%%%%%%%%
%%%%%%%%%%%%%%%%%%%%%%%%
\newpage

%\section{Supplementary information}

%%%%%

\bibliographystyle{ieeetr}
\bibliography{networksbibliography}

\end{document}